\documentclass[a4paper,fleqn]{cas-dc}

\usepackage[authoryear,longnamesfirst]{natbib}
\usepackage{caption}
\usepackage{multirow}
\usepackage{makecell, cellspace, caption}
\usepackage{balance}
\usepackage{epstopdf}
\usepackage{siunitx}

\def\tsc#1{\csdef{#1}{\textsc{\lowercase{#1}}\xspace}}
\tsc{WGM}
\tsc{QE}
\tsc{EP}
\tsc{PMS}
\tsc{BEC}
\tsc{DE}

\begin{document}
\let\WriteBookmarks\relax
\def\floatpagepagefraction{1}
\def\textpagefraction{.001}

\shorttitle{}

\shortauthors{Kawshik Nath et~al.}

\title [mode = title]{High-Efficiency Hexagonal Nanowire MAPbI$_3$ Perovskite Solar Cell with Broadband Light Trapping}



%
\author[1,2]{Kawshik Nath}[auid=000,bioid=1,
                        orcid=0009-0001-9749-278X]





\credit{Conceptualization, Methodology, Visualization, Software, Investigation, Writing – original draft}

\affiliation[1]{organization={ Department of    Electrical and Electronic Engineering,        Bangladesh University of Engineering and      Technology}, 
    city={Dhaka},
    country={Bangladesh}}
    \affiliation[2]{organization={Department of Electrical and Electronic Engineering, Chittagong University of Engineering and Technology},
    city={Chattogram},
    country={Bangladesh}}
    
    \author[1,2]{Bibekananda Nath}[auid=000,bioid=1,
                        orcid=0009-0000-2213-6628]





\credit{Conceptualization of this study, Methodology, Software}


\author[1]{Ahmed Zubair}[orcid=0000-0002-1833-2244]
\cormark[1]
\credit{Conceptualization of this study, Methodology, Software}
\ead{ahmedzubair@eee.buet.ac.bd}
\ead[URL]{http://ahmedzubair.buet.ac.bd/}





\cortext[cor1]{Corresponding author}



\begin{abstract}
Perovskite solar cells (PSCs) have emerged as strong contenders for the next generation of photovoltaic (PV) technologies due to their exceptional light absorption properties, tunability, and affordability in manufacturing. Here, we presented an ingenious hexagonal nanowire (HNW)-based PSC that achieves broadband absorption, minimizes reflectance, and offers robust polarization insensitivity by improving light-matter interaction and increasing charge-collection efficiency. The rotational symmetry of the HNW configuration yielded polarization-independent absorbance under both TE and TM illumination across the visible and near-infrared spectra. The optimization of the geometrical parameters of  CH$_3$NH$_3$PbI$_3$-based HNW  structure, including diameter, period, and fill ratio, offered a wide range of variations that influenced both optical properties and device performance. To further intensify photon confinement, a dielectric SiO$_2$ sphere is partially embedded in the ITO layer, improving long-wavelength absorbance and increasing electron–hole pair generation near the active region. We analyzed the finite-difference time-domain (FDTD) method to examine the optical properties of our proposed structure. This study demonstrates that our proposed structure has achieved a higher generation rate, enhanced absorbance, and a higher optical short-circuit current density ($J_{\text{sc}}$) of 29.53 mA/cm$^2$. Electrical performance is assessed by solving the coupled drift-diffusion and Poisson equations for the dynamics of carrier transport. The optimized HNW structure achieved a notable power conversion efficiency of 24.2\%, highlighting a strong connection between optical confinement and effective carrier transport. These attributes render the proposed HNW PSC a viable option for high-performance PV systems and scalable thin-film solar technologies.

\end{abstract}





\maketitle

\section{Introduction}
To address the escalating energy demands of the 21\textsuperscript{st} century, solar cell technologies have been increasingly directed toward the efficient generation of renewable solar power, demonstrating rapid and sustained growth in recent years \cite{D4TA06224H,rahman2025,Akhtary:23,D3NA00565H}. Among the various photovoltaic (PV) technologies, perovskite solar cells (PSCs) have emerged as leading contenders due to their tunable bandgaps, extended carrier diffusion lengths, enhanced charge transport properties, and straightforward fabrication processes. Besides, PSCs have rapidly gained recognition as one of the most promising PV technologies, offering the potential for high efficiency at reduced production costs~\cite{Han2025,pandey2023halide,al2023understanding}. Among different organic materials, MAPbI\textsubscript{3} has been widely adopted in PSCs since its initial introduction by Kojima \textit{et al.} in 2009 \cite{kojima2009organometal}. Furthermore, comprehensive studies demonstrated that lead (Pb)-based PSCs exhibit superior power conversion efficiencies (PCEs) compared to their lead-free counterparts \cite{shao2018highly, Sharma2023, Faiaad2026}. Both industrial and academic communities have shown significant interest in the emerging technology of lead (Pb) halide PSCs, particularly due to their remarkable advancements in PCEs, which surpassed 23 \% \cite{jeon2018fluorene}. Additionally, the open-circuit voltages ($V\textsubscript{oc}$) of perovskite devices are approaching the theoretical radiative limit, for instance, 1.32 V of $V\textsubscript{oc}$ was reported for MAPbI\textsubscript{3}\,\cite{tress2015predicting, rau2017efficiency, correa2017promises}, given its bandgap ($E_\text{g} \approx 1.6 eV$)~\cite{sharma2023numerical}. The majority of solar cell architectures employ either mesoporous or planar configurations, incorporating a perovskite absorber layer sandwiched between the hole-transport layer (HTL) and the electron-transport layer (ETL).

Recently, Wang \textit{et al.} developed TiO\textsubscript{2}/MAPbI\textsubscript{3}/NiO PSCs with TiO\textsubscript{2} nanowire (NW) ETL arrays and reported a PV conversion efficiency of 21.8\% \cite{wang2024effects}. Hmaidi  \textit{et al.} proposed Si NW as HTL for MAPbI\textsubscript{3} perovskite material and found 11.93\% efficiency \cite{hmaidi2025pedot}. To improve the absorbance range (200-2500 nm) and eventually the efficiency of the PSC, Ren\textit{ et al.} combined quantum dots (QDs) with perovskite material and created an ETL/QD/perovskite layer/HTL core-shell structure NW array solar cell \cite{ren2024new}. Moreover, Liang \textit{et al.} designed an Ag NW-based MAPbI\textsubscript{3} PSC to achieve tunable transmittance, reporting a maximum efficiency of 17.3\% \cite{wang2025efficient}. To further enhance performance, they designed a homojunction p-type MAPbI\textsubscript{3}/n-type MAPbI\textsubscript{3} solar cell to reduce the generated carriers' recombination by creating an internal electric field, and they achieved an efficiency of 22\% for a certain thickness of the absorber layer \cite{liang2024construction}. To push the PCE beyond 25\%, recent research efforts have focused on advanced architectures such as bifacial 3D/3D, 3D/2D, or even multi-stacked perovskite layers. Alathlwi \textit{et al.} proposed a PET/Ag NWs: MXenes/SrCuO\textsubscript{2}/
\mbox{(FAPbI\textsubscript{3})\textsubscript{0.95}(MAPbBr\textsubscript{3})\textsubscript{0.05}}/C\textsubscript{60}/BCP/FTO solar cell structure, achieving a high PCE of approximately 26\% \cite{alathlawi2025role}. Mkwai \textit{et al.} fabricated a GeS-doped MAPbI\textsubscript{3} solar cell and reported an efficiency of 17.46\% \cite{mkawi2024enhancement}. However, 2D/3D structures primarily suffer from high series resistance and poor charge transport efficiency due to unmatched lattice parameters and non-symmetrical conductivity. To overcome such issues, Patil \textit{et al.} proposed a 3D/3D bilayer MAPbI\textsubscript{3}/FAPbI\textsubscript{3} heterojunction solar cell and obtained an PCE of 23.08\% \cite{patil2025exceeding}. However, these types of multilayer structures are both costly and complex to fabricate. To overcome such issues, several studies have developed monolayer perovskite absorbers of various shapes. Im \textit{et al.} fabricated NW-shaped MAPbI\textsubscript{3} through the two-step spin-coating procedure and reported an efficiency of 14.71\% under standard AM 1.5G solar irradiation \cite{im2015nanowire}. Sing \textit{et al.} further improved the efficiency of the NW perovskite to 18.7\% by adding fullerene derivative (PC\textsubscript{60}BM) \cite{singh2018enhancing}. To eradicate the ionic defects at the grain boundary and from the surface to improve the efficiency of organic-inorganic PSCs, Cha \textit{et al.} demonstrated a nanowire hybrid PSC (NWHPSC) and charge transport layers with an efficiency of 21.56\% and stability up to 3500 hours \cite{cha2023perovskite}. Despite significant advancements in PSCs utilizing multilayer heterostructures, 2D/3D junctions, and nanowire morphologies, these configurations often encounter issues such as polarization-dependent optical responses, inefficient electromagnetic field localization, and the need for additional transport layers. Moreover, conventional NW architectures generally enhance absorbance via localized resonances but
lack geometrical symmetry, resulting in a nonuniform field localization and minimal carrier extraction. Moreover, the inefficient field localization hinders the capability of broadband absorption under the unpolarized illumination. There is a scope for research to design structures that mitigate these limitations by enhancing light-matter interaction, near-field confinement, and efficient carrier extraction within a single absorber framework \cite{NATH2026107697}.

In this work, we conducted a comprehensive optoelectronic study of an Hexagonal Nanowire (HNW)-based CH\textsubscript{3}NH\textsubscript{3}PbI\textsubscript{3} (MAPbI\textsubscript{3}) PSC aimed at achieving efficient light trapping by employing physical mechanisms governed by the light matter interaction and the carrier dynamics. We investigated the optical and electrical performance of the proposed structure using the finite-difference time-domain (FDTD) method, coupled with a self-consistent solution of drift-diffusion and the Poisson equation for the transverse electric (TE), transverse magnetic (TM), and unpolarized light  \cite{ANSYS_Lumerical}. Our findings indicated that the HNW design facilitates a significant enhancement in broadband absorption and effective charge extraction. It showed behavior that is independent of polarization by enhancing the near-field confinement. 
Additionally, an analysis of the generation rate profile and electric field distributions revealed the influence of the hexagonal architecture in directing and localizing light, which in turn enhanced carrier generation and the overall PV performance.  The electrical simulations demonstrated a high PCE attributed to the broadband light absorbance throughout our proposed structure and exhibited superior performance compared to the recently developed MAPbI\textsubscript{3} based PSCs of various configurations. Alongside the numerical analysis, this study addresses the practical challenges of experimentally realizing the structures to enable their commercial utilization.

\begin{figure*}[h]
  \centering
   \includegraphics [width=1\textwidth]{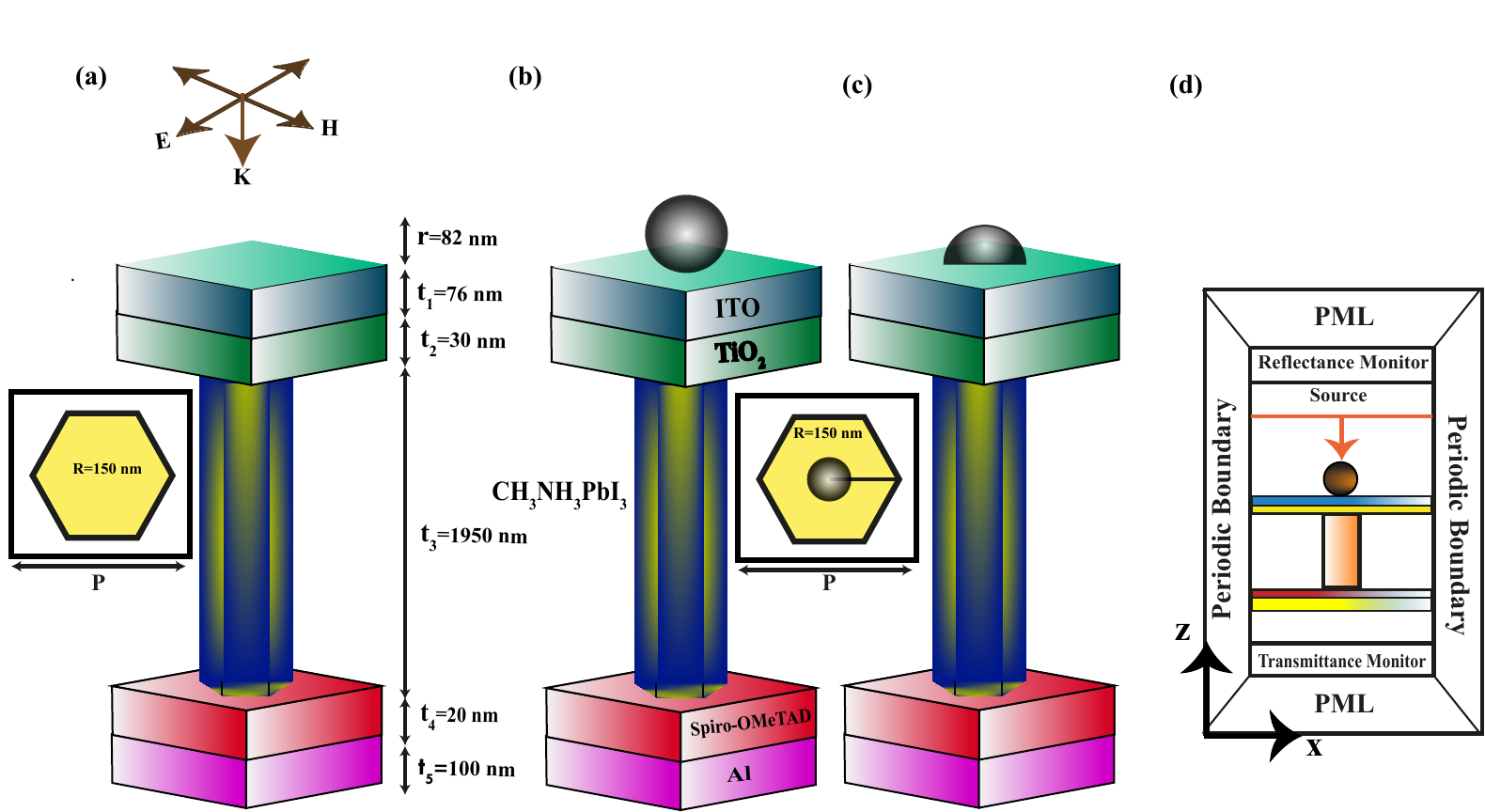}
    \caption{(a) Schematics of the unit cell of an HNW-based nanostructured PSC, (b) including a SiO$_2$ based sphere on top of the ITO, (c) inserting the sphere inside the ITO, (d) cross-sectional view of the xz plane of our proposed structure with simulation setup.}
    \label{Fig:fig1}
\end{figure*}

\section{Structure Design and Simulation Methodology}
\subsection{Structural Design}
The proposed structure consists of an HNW-based PSC as shown in Fig.\,\ref{Fig:fig1} (a). The design includes a vertically aligned HNW made of MAPbI\textsubscript{3} perovskite material. To improve the light confinement and scattering along the absorber region, a SiO$_2$ based sphere is introduced in two different configurations: one placed on top of the indium-doped tin oxide (ITO) layer and another one partially embedded inside it according to Figs.\,\ref{Fig:fig1} (b), and (c). 
This configuration allows for Mie resonance and enhanced light confinement. The optimal thickness of each layer is also given in this figure.
Here, the dielectric sphere acts as an optical concentrator, directing the incident photons into the perovskite layer by enhancing light coupling, promoting scattering, and minimizing reflectance losses. The ITO layer simultaneously allows for high optical transmittance and efficient lateral charge transport while exhibiting negligible parasitic absorbance across the visible spectral range. The ETL made of TiO$_2$ allows for the selective extraction of electrons from the photoactive perovskite layer and guides the electrons toward the metal contact, effectively preventing hole recombination through proper energy band alignment. Conversely, HTL made of spiro-OMeTAD promotes the efficient collection of generated holes while also serving as an electronic barrier to prevent unwanted electron backflow, thus ensuring strong charge carrier selectivity. 
The rear metal contact enables the removal of stored charge carriers and completes the PV circuit with minimal resistive and optical losses. In this study, we used aluminum (Al) as a back contact.

\subsection{Simulation Methodology}
We conducted optical simulations of the proposed structure utilizing the three-dimensional FDTD method using \emph{Ansys Lumerical} \cite{ANSYS_Lumerical}. Due to the periodicity of the structure along the $x$ and $y$ axes, we performed the simulation over a unit cell and applied periodic boundary conditions in both directions. To reduce unwanted reflectances along the $z$ axis, we incorporated 64 perfectly matched layers (PMLs) on both the top and bottom of the model. We chose a non-uniform mesh with a step size of \SI{0.25} {nm}, and also incorporated an override mesh for each layer to ensure better accuracy. Mesh refinement was performed by implementing the integral solution of Maxwell's equations near the interface for all the materials, including metals, which can lead to better convergence and accuracy.

We utilized a downward-directed continuous wave (CW) normalized plane wave source, spanning wavelengths from 300 to 850 nm, to analyze the performance of our designed perovskite cell for TM, TE, and unpolarized incident waves. To determine the optical performance, frequency-domain field and power monitors were employed to enumerate the transmitted and reflected light passing through the photoactive layer. These monitors captured the characteristics of transmittance, $T(\lambda)$, and reflectance, $R(\lambda)$, spectra as a function of wavelength. The simulation environment is illustrated in Fig.\,\ref{Fig:fig1} (d). To achieve accurate optical simulations, the refractive index ($n$) and extinction coefficient ($k$) for all materials were carefully incorporated into the model. The dispersive optical properties for Al and SiO$_2$ were collected from Palik \textit{et al.} \cite{palik1998handbook} and for  TiO$_2$, Spiro-OMeTAD, and MAPbI\textsubscript{3} were gathered from Elewa \textit{et al} \cite{Elewa}.

The transmittance and reflectance spectra are calculated by,

\begin{equation}
    R(\lambda) = \frac{P_R}{P_I}, 
\end{equation}
\begin{equation}
    T(\lambda) = \frac{P_T}{P_I}.
\end{equation}
Here, $P_I$, $P_R$, and $P_T$ indicate the incident, reflected, and transmitted wave's power, respectively. The absorbance, $A(\lambda)$ is wavelength-dependent and can be determined by, 
\begin{equation}
    A(\lambda) = 1-(R(\lambda) + T(\lambda)).
\end{equation}

The electrical performance parameters of the proposed PSC were determined by using Poisson's and drift-diffusion equations as follows:
\begin{align}
J_n &= q\mu_n n E + q D_n \nabla n,\\
J_p &= q\mu_p p E - q D_p \nabla p.
\end{align}

Here, \( J_n \) and \( J_p \) indicate the current densities of electrons and holes, respectively, measured in $\mathrm{mA\,cm^{-2}}$. The parameter \( q \) stands for the charge of an electron, while \( \mu_n \) and \( \mu_p \) refer to the mobilities of electrons and holes. The electric field is represented by \( E \), and \( D_n \) and \( D_p \) denote the diffusion constants of electrons and holes, respectively. The concentrations of charge carriers for electrons and holes are represented by \( n \) and \( p \), respectively. 
The characteristics of these charge carriers are determined by two main mechanisms: drift, which is affected by an external electric field, and diffusion, which occurs due to concentration differences driven by thermal energy. Carrier mobility \( (\mu_p) \) quantifies the speed at which electrons or holes move through a semiconductor when an electric field is present. It is a crucial factor in assessing the efficiency of charge transport. This mobility is also thermodynamically connected to the diffusivities \( D_n \) and \( D_p \) through the Einstein relation, which is given by,
\begin{equation}
D_{n(p)} = \mu_{n(p)} \frac{k_B T_{abs}}{q}
\end{equation}

Here, \( k_B \) represents the Boltzmann constant, \( T_{abs} \) denotes the absolute temperature, and \( q \) indicates the charge of the carriers. This relationship provides a theoretical basis that links the random thermal motion of carriers (diffusion) with their directed movement under an electric field (drift), thereby connecting microscopic thermal dynamics to the macroscopic electrical properties of semiconducting materials.
\begin{figure*}[t]
  \centering
   \includegraphics {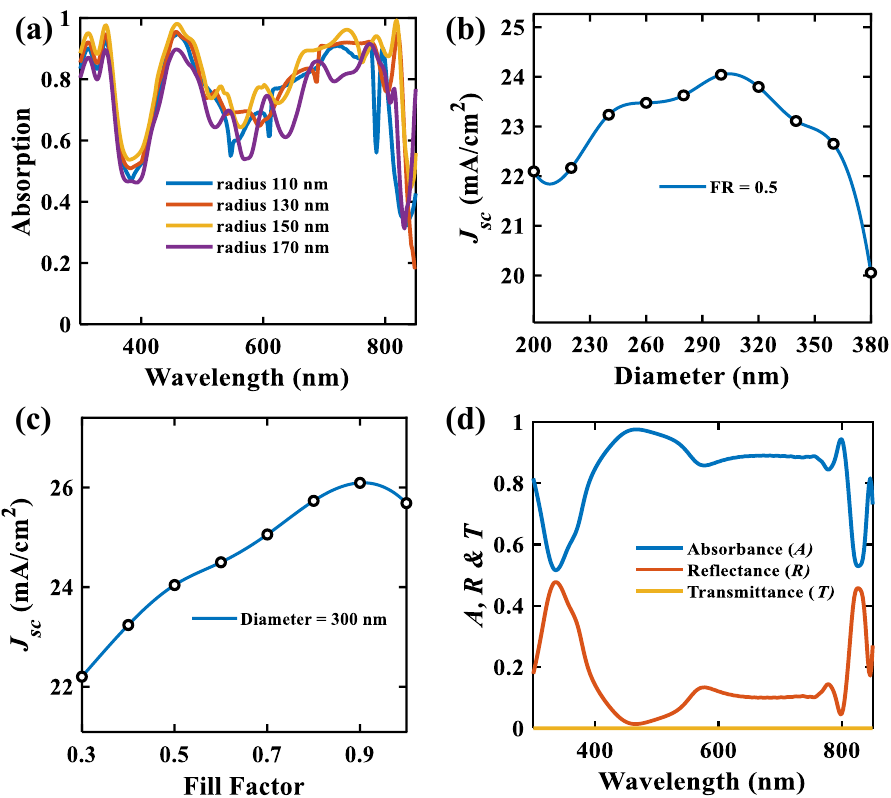}
    \caption{(a) Absorbance spectra for different radius at a fixed FR of 0.5, (b) optical $J_{\text{sc}}$ for different radius at FR = 0.5, (c) optical $J_{\text{sc}}$ for a fixed radius of 150 nm for varying FR, (d) absorbance, reflectance, and transmittance spectra for FR = 0.9 at a radius of 150 nm.}

    \label{Fig:fig2}
\end{figure*}
The Shockley–Read–Hall (SRH) mechanism is the major non-radiative recombination process, which significantly limits the efficiency of PSCs. As a result, the kinetics of carrier recombination in the perovskite absorber layer are analyzed quantitatively using the expression for the SRH recombination rate.
\begin{equation}
R_{\mathrm{SRH}} = \frac{np - n_i^2}{\tau_n (n + n_i) + \tau_p (p + n_i)}
\end{equation}

Here, \( \tau_n \), \( \tau_p \), and \( n_i \) are the lifetimes of the electron and the hole, and the intrinsic carrier concentration, respectively.

The total number of photons absorbed can be calculated by taking the absorbance profile \( P_{\mathrm{abs}} \) and dividing it by the energy corresponding to a single photon, represented by the following equation:
\begin{equation}
G = \frac{P_{\mathrm{abs}}}{\hbar \nu}
\end{equation}

Here, \( \hbar \) stands for the reduced Planck’s constant and \( \nu \) is the frequency of light. The generation rate \( G \) is derived by integrating this photon flux across the relevant wavelength or frequency range within the simulated spectrum.

Assuming unit internal quantum efficiency, the resulting photogeneration current \( I_p \) can be approximated as equation \ref{eq9} by considering each absorbed photon results in the generation of an electron–hole pair:
\begin{equation}
I_p = eG
\label{eq9}
\end{equation}

The quantum efficiency \( \mathrm{QE}(\lambda) \) of a solar cell can be expressed by,
\begin{equation}
\mathrm{QE}(\lambda) = \frac{P_{\mathrm{abs}}(\lambda)}{P_{\mathrm{in}}(\lambda)}
\end{equation}

Here, \( P_{\mathrm{in}}(\lambda) \) and \( P_{\mathrm{abs}}(\lambda) \) represent the incident and absorbed radiative powers into the PSC at a particular wavelength \( \lambda \), respectively.

The impact of geometric features, such as size and shape, on device performance was then evaluated by analyzing power–voltage (P–V) and current density–voltage (J–V) characteristics. 
The short-circuit current density ($ J_{\mathrm{sc}}$) was calculated by the following equation:
\begin{equation}
J_{\mathrm{sc}} = \frac{q}{\hbar c} \int_{300\,\mathrm{nm}}^{850\,\mathrm{nm}} \mathrm{EQE}(\lambda) \, \lambda \, I_{\mathrm{AM1.5}}(\lambda) \, d\lambda
\end{equation}

Here, \( c \) refers to the speed of light. The upper limit of the integration corresponds to the wavelength determined by the bandgap of the perovskite material, which prevents absorbance below this energy threshold, as absorbance of photons below this will not contribute to electron-hole pair generation.

The PCE is another important metric used to examine the electrical performance of solar cells and is defined by,
\begin{equation}
\eta = \frac{\mathrm{FF} \cdot V_{\mathrm{oc}} \cdot J_{\mathrm{sc}}}{P_{\mathrm{AM1.5G}}}.
\end{equation}

\( P_{\mathrm{AM1.5G}} \) represents the incident solar power density from the AM1.5G spectrum, usually taken as 100 {mW/cm\textsuperscript{2}}.

\section{Results and Discussion}
\subsection{ Optical Analysis}
\subsubsection{Geometric Optimization for Light Management in HNW}
The optical properties of HNW solar cells are significantly influenced by their geometric parameters, including diameter ($D$), period ($P$), and the fill ratio ($FR$), which can be defined as $D/P$. The objective of the optical simulation was primarily to maximize the photocurrent generated by absorbed photons, which also relies on the geometric parameters of the nanostructures. Fig.\,\ref{Fig:fig2}(a) depicts the absorbance spectra for different NW diameters ($D$) maintained at a constant fill ratio ($FR = 0.5$). As the radius ($R = D/2$) decreases from 110~nm to 150~nm, the absorbance characteristics become broader and more uniform in the 450-800~nm spectrum. This improvement stems from enhanced light confinement, increased photonic coupling, and longer effective optical paths within the perovskite NW, leading to higher photon absorbance. The peak optical $J_{\text{sc}}$ of 24.01~mA/cm$^2$ is observed at $R = 150$~nm, from Fig.\,\ref{Fig:fig2}(b). 
At $R = 170$~nm, the absorbance curve experiences a decline, especially in the 600-800~nm range, where longer-wavelength photons are not efficiently captured. Due to the more efficient excitation and confinement of the photons inside the NW, HNWs with a diameter of 300~nm demonstrated significantly better optical absorbance~\cite{im2015nanowire}. Therefore, further increasing the NWs radius beyond 150~nm degrades the absorbance, as can be observed from Fig.\,\ref{Fig:fig2}(a).

 After the optimized diameter for HNW was determined, the periodicity of the NW needs to be optimized to achieve the most efficient and well-structured design. To achieve this, NW's with a constant radius of 150 nm were simulated at various $FR$s ranging from 0.3 to 1.0. In Fig.\,\ref{Fig:fig2}(c), the photon-generated current density is shown for a fixed radius. Due to the inverse relationship between the $FR$ and $P$, a lower $FR$ corresponds to a wider interwire spacing. Structures with $FR~$<$~0.5$ exhibit inadequate absorbance, attributed to weaker field localization and reduced optical confinement, as indicated by the simulation results. Conversely, $FR$ values in the range of 0.7 to 0.9 demonstrate improved $J_{\text{sc}}$ and enhanced light absorbance, indicating a more favorable photonic coupling. The optimum configuration yields a peak optical $J_{\text{sc}}$ of 26.09~mA/cm$^2$, corresponding to a NW diameter of 300~nm and an $FR$ of 0.9. This optimized structure represents a well-balanced design that maximizes both carrier generation and light-matter interaction in the HNW array. Subsequently, for this proposed configuration,  the ETL and HTL layers were optimized, resulting in a peak optical $J_{\text{sc}}$ of 27.02~mA/cm$^2$.   Fig.\,\ref{Fig:fig2}(d) represents the absorbance, reflectance, and transmittance spectra for the fill ratio of 0.9 and a diameter of 300 nm. From this figure, the absorbance curve (blue) indicates consistently high values greater than 85\% throughout most of the visible spectrum, particularly between 450 and 780 nm, which reveals an outstanding ability to harvest light. The high level of absorbance is due to the strong interaction between light and matter, facilitated by the hexagonal structure, which improves optical confinement. From the transmittance spectra, it can be seen that the transmittance is nearly zero, which means most of the light is absorbed by the structure and a small portion is reflected.
 \begin{table*}[htbp]
\centering\caption{Electrical properties of the materials used in this simulation}
\small
\begin{tabular}{lccc}
\toprule
\textbf{Property} & \textbf{MAPbI$_3$ (Perovskite)} & \textbf{TiO$_2$ (ETL)} & \textbf{Spiro-OMeTAD (HTL)} \\
\midrule
DC Permittivity, $\varepsilon$ & 6.25 & 7.84 & 7.5 \\
Work Function (eV)              & 4.68 & 5.6 & 3.655 \\
$E_c$ Valley                    & X & X & X \\
Effective mass, $m_n$ (1/$m_e$) & 4.63 & 11.67 & 4.63 \\
Effective mass, $m_p$ (1/$m_e$) & 4.63 & 3.99 & 4.63 \\
Bandgap, $E_g$ (eV)             & 1.5 & 3.2 & 2.91 \\
Acceptor doping, $N_A$ (cm$^{-3}$) & $2.4 \times 10^{17}$
 & -- & $3 \times 10^{18}$
 \\
Donor doping, $N_D$ (cm$^{-3}$) & -- & $5 \times 10^{19}$ & -- \\
Electron mobility, $\mu_n$ (cm$^2$/Vs) & 50 & 0.006 & 0.0001 \\
Hole mobility, $\mu_p$ (cm$^2$/Vs) & 50 & 0.006 & 0.0001 \\
Auger Recombination coefficient for electron, $C_n$ (cm$^6$/s) & $2.7 \times 10^{-29}$ & -- & -- \\
Auger Recombination coefficient for hole, $C_p$ (cm$^6$/s) & $4.6 \times 10^{-29}$ & -- & -- \\
Trap Assisted Recombination coefficient for electron, $\tau_n$ (ns) & 8 & 5 & 5 \\
Trap Assisted Recombination coefficient for hole, $\tau_p$ (ns) & 8 & 5 & 2 \\
\bottomrule
\end{tabular}
\label{tab:material_properties}
\end{table*}

\begin{figure}[]
  \centering
   \includegraphics{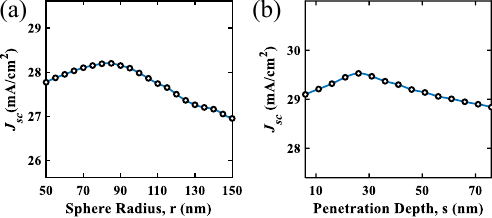}
    \caption{ Optical $J_{\text{sc}}$ values (a) SiO\textsubscript{2} sphere on ITO layer, (b) SiO\textsubscript{2} sphere partially embedded into the ITO layer.}

    \label{Fig:fig3}
\end{figure}

\begin{figure*}[t]
  \centering
   \includegraphics {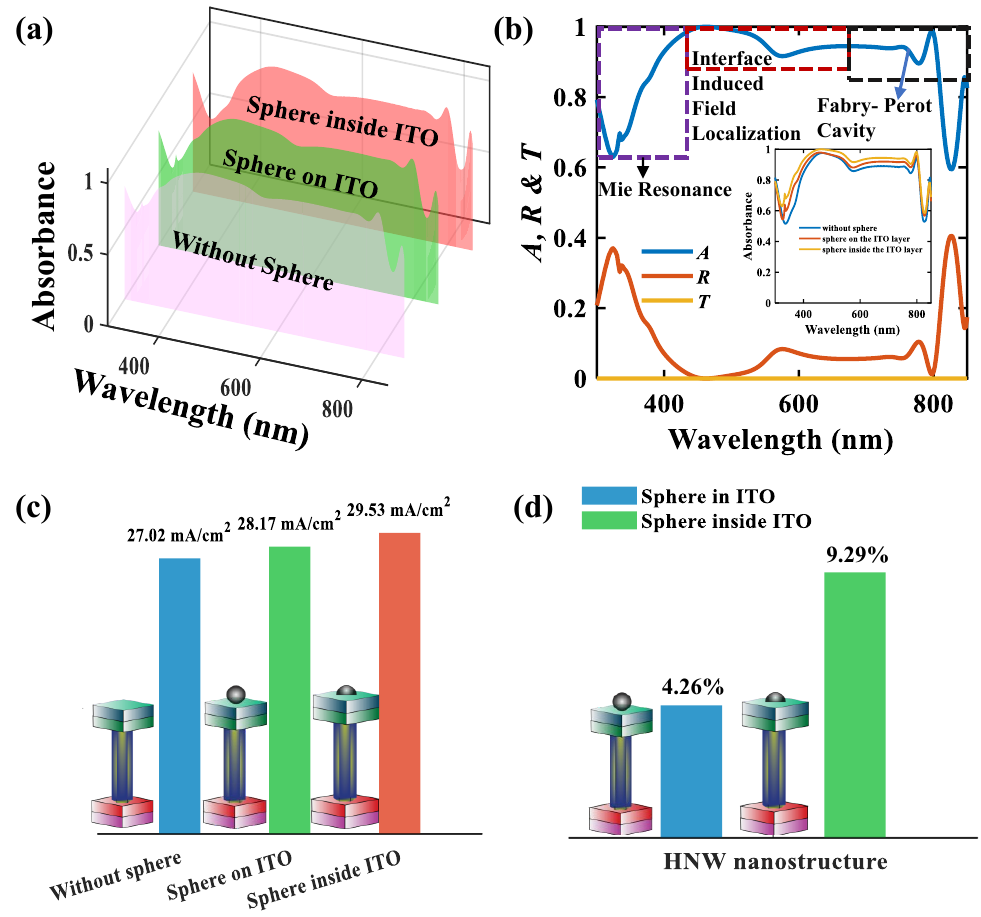}
    \caption{(a) Absorbance spectral density for three different configurations of the optimized structure. The inset shows the variation of absorbance spectra for the three configurations, (b) absorbance, reflectance, and transmittance spectra for the configuration of penetrating the sphere into the ITO layer, (c) $J_{\text{sc}}$ values for these three different configurations (d) percentage (\%) improvement in $J_{\text{sc}}$.}

    \label{Fig:fig4}
\end{figure*}

To further enhance light confinement and improve device performance beyond the optimized HNW configuration, we performed a radius-dependent optimization study by incorporating a dielectric SiO$_2$ sphere on top of the ITO layer. The SiO$_2$ sphere was introduced with two primary objectives. First, it promotes stronger light trapping by encouraging constructive interference and enhancing the coupling of incoming photons due to the enhanced field localization of the HNW array. Second, it introduces negligible parasitic absorbance in the operating wavelength range \cite{LUO2018128}. From this analysis, we achieved the maximum optical $J_{\mathrm{sc}}$ of 28.2 mA/cm$^2$ at a radius ($r$) of 82 nm identified a $r$ of 82\,nm as demonstrated in Fig.\,\ref{Fig:fig3} (a). Guided by this result, the optimized SiO$_2$ sphere was subsequently integrated into the structure, allowing it to partially penetrate the ITO layer and enabling the effective redirection and redistribution of incoming light into the perovskite NW region. This design modification scatters and redirects incoming light into the active region, thereby increasing the optical path length within the nanostructured absorber. Consequently, the addition of sthe sphere mitigates the reduced absorbance observed at higher filling ratios and results in improved broadband absorption and an increased optical $J_{\mathrm{sc}}$ of 29.53~mA/cm$^{2}$ as depicted in Fig.\,\ref{Fig:fig3} (b). Here, we considered the penetration depth as 's' in~nm.

This improvement is further validated by Fig.\,\ref{Fig:fig4}(a), which represents the absorbance spectra for three different configurations: (i) HNW structure without SiO$_2$ sphere, (ii) SiO$_2$ sphere on ITO layer, and (iii) SiO$_2$ sphere partially embedded inside ITO layer. The figure shows that when the SiO$_2$ sphere is embedded within the ITO layer, absorbance exceeding 90\% is sustained across the 400--800~nm wavelength range. Fig.\,\ref{Fig:fig4}(b) shows the simulated spectra for absorbance (A), reflectance (R), and transmittance (T) of the optimized HNW PSC when the SiO$_2$ sphere is in the ITO layer. Two distinct absorbance peaks are observed. A notable increase in absorbance around 430 nm to 650 nm is linked to the coupling of high-energy photons via pronounced interface-induced field localization between successive layers. A second significant peak in the range of 830–850 nm is probably due to the Fabry-Pérot (FP) type interference occurring at longer wavelengths. The high level of absorbance is attributed to the improvement of optical confinement and effective coupling of incoming light into Mie-resonant and FP-cavities present within the absorber. 

In order to provide a more comprehensive physical understanding of the observed absorbance features, we utilized analytical expressions derived from classical electrodynamics and Mie scattering theory, which demonstrate how the sphere enhances the light matter interaction into the absorber layer \cite{bohren2008absorption,novotny2012principles}. The fundamental descriptor governing the sphere–light interaction is the Mie size parameter,
\begin{equation}
    x = \frac{2\pi r}{\lambda},
\end{equation}
which determines the shift between Rayleigh ($x\ll1$), Mie ($x\sim1$), and geometric-optics ($x\gg1$) regimes. For the optimized sphere radius of $r=82~\mathrm{nm}$, the size parameter varies from $x=0.60$ to $1.47$ over the operating wavelength range ($350$–$850$ nm), placing the structure firmly in the resonant Mie-scattering regime where strong forward scattering and significant near-field amplification occur.

Under these conditions, the scattering efficiency is governed by the following equation \cite{novotny2012principles},
\begin{equation}
    Q_{\mathrm{sca}} = \frac{2}{x^2}\sum_{n=1}^{\infty}(2n+1)\left(|a_n|^2 + |b_n|^2\right),
\end{equation}
where $a_n$ and $b_n$ are the electric and magnetic Mie coefficients, respectively. 
This equation governs that moderate x-values enhance multipolar resonance excitation while minimizing the reflected power. On the other hand, due to the negligible parasitic absorbance of SiO$_2$, it redistributes upcoming photons into the absorber region. This redistribution increases the local optical field intensity according to the equation,
\begin{equation}
    p_{\mathrm{abs}} = \frac{1}{2}\,\omega \varepsilon_0\, \mathrm{Im}[\varepsilon]\; |E|^2,
\end{equation}

Figs.\,\ref{Fig:fig4}(c) and (d) represent the optical $J_{\text{sc}}$ and percentage (\%) improvement, $J_{\text{sc}}$ and from these figures it is clearly seen that maximum $J_{\text{sc}}$ is found in the configuration three for higher optical confinement in the nanostructure.  
\begin{figure*}[t]
  \centering
   \includegraphics
   [width=1.0\textwidth]{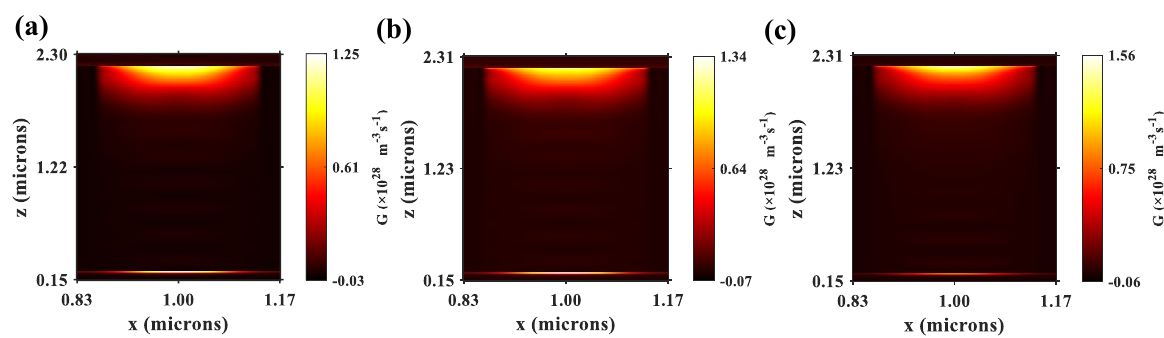}
    \caption{Generation profile for our proposed structure: (a) without sphere, (b) sphere on top of ITO layer, (c) sphere inside the ITO layer.} 
    \label{Fig:fig5}
\end{figure*}
\begin{figure*}[t]
  \centering
   \includegraphics
   [width=1.0\textwidth]
   {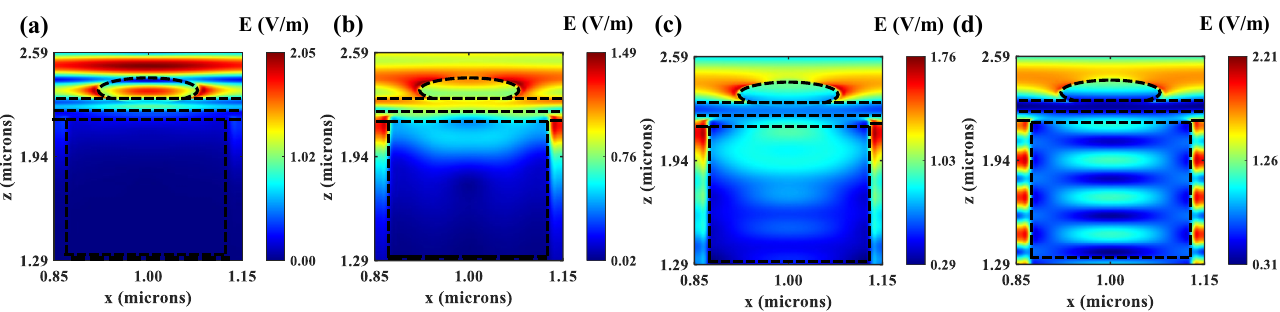}
    \caption{Electric field distribution at different wavelengths: 
(a) $\lambda = 300$~nm, 
(b) $\lambda = 430$~nm,  
(c) $\lambda = 762$~nm, and 
(d) $\lambda = 826$~nm for the structure with SiO$_2$ sphere inside the ITO layer.}
\label{Fig:fig6}
\end{figure*}
Moreover, this improvement is supported by the generation profiles for these three configurations. According to Figs.~\ref{Fig:fig5}(a)--(c), our proposed structure exhibits a higher optical ($J_{\text{sc}}$) of 29.53~mA/cm$^2$ and an enhanced generation rate ( $1.56 \times 10^{28}\,\mathrm{m^3/s}$
) across the absorber layer. The configuration with the SiO$_2$ sphere embedded inside the ITO layer shows the most efficient light trapping and photogeneration, confirming the effectiveness of this design in improving the performance of the device. The high generation rate is directly linked to the robust electric field confinement present in the active region, as demonstrated by the electric field profile detailed below.

Fig.\,\ref{Fig:fig6} (a) illustrates the electric field distribution at a wavelength of 300 nm. The field profile exhibits significant confinement of the photon at the upper hemispherical section of the SiO$_2$ sphere, while minimal field penetration occurs into the perovskite layer. This phenomenon characterizes the high-order Mie resonance that arises from the refractive index difference between the SiO$_2$ sphere and its surrounding medium. Despite the strong intensity of the field at the sphere surface, the limited field localization into the perovskite layer leads to insufficient light harvesting at this wavelength due to the low optical penetration depth. In Fig.\,\ref{Fig:fig6} (b), at a wavelength of 430 nm, the electric field starts to spread more effectively toward the lower half of the SiO$_2$ sphere and the upper part of the perovskite layer. This can be due to the interface-induced field confinement followed by low Fresnel reflection rather than any explicit resonant mode. Here, the SiO$_2$ sphere changes the direction of the incident plane waves into near-field energy. As a result, the HNW perovskite layer begins to support broadband absorption, which increases the intensity of the internal field. 
At 762 nm, as depicted in Fig.\,\ref{Fig:fig6} (c), the electric field exhibits clear vertical guiding patterns within the HNW. This indicates the FP-cavity effect due to the internal reflectance and disperses the propagation of the localized field across the device due to the symmetry of the HNW. At 826 nm, as illustrated in Fig.\,\ref{Fig:fig6} (d), a standing wave pattern has been observed throughout the HNW.   On the other hand, the SiO$_2$ sphere acts as a passive coupling layer due to its lower refractive index compared to the surrounding materials at this wavelength regime.

\begin{figure}[]
  \centering
   \includegraphics[scale=0.9]{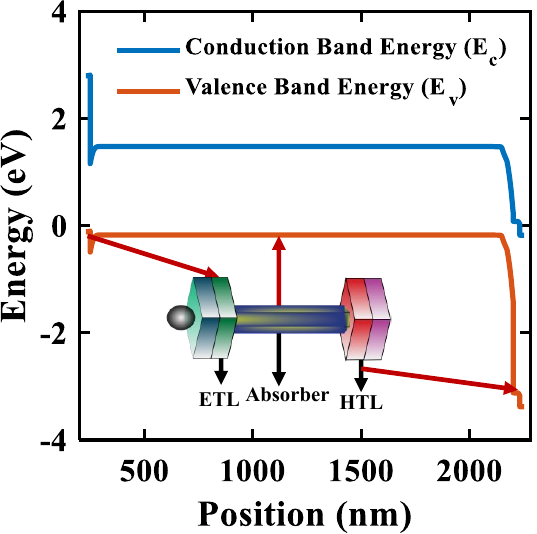}
    \caption{ Band alignment across the layer of the optimized HNW PSC.}

    \label{Fig:fig7}
\end{figure}

Furthermore, to gain a better understanding of the carrier transport mechanisms in our proposed PSCs, the energy band alignment across the HTL to ETL is shown in Fig.\,\ref{Fig:fig7}. The conduction band (CB) and valence band (VB) levels of each layer are selected to facilitate efficient charge separation and directional transport. From the figure, it can be observed that there is a smooth transition of the charge carriers throughout the cell, indicating effective carrier transport throughout the device. This smooth band alignment across the interfaces minimizes potential barriers, suppressed recombination, and supports a strong built-in electric field. This plays a significant role in achieving high $J\textsubscript{sc}$ and $V\textsubscript{oc}$, ultimately enhancing the overall PCE.

\begin{figure*}[t]
  \centering
   \includegraphics 
   [width=0.98\textwidth]{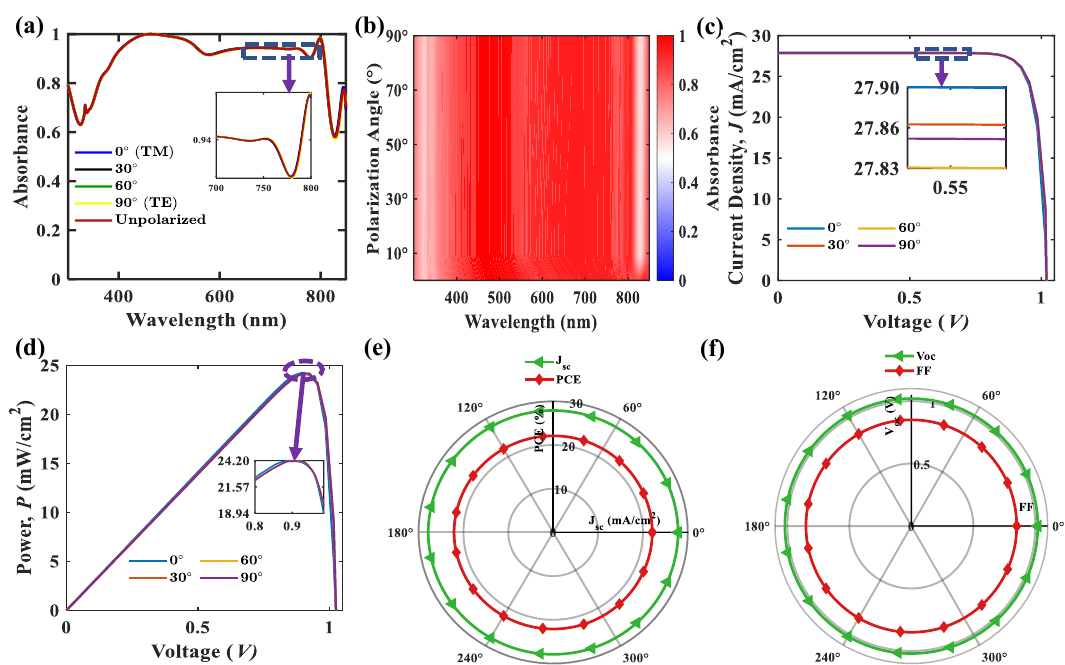}
\caption{(a) Absorbance spectra for different polarization states and an unpolarized light source of the optimized structure,
(b) color map of polarization angle-dependent absorbance spectra, 
(c) and (d) $J$--$V$ and $P$-$V$ curves for the proposed HNW PSC; the inset shows the zoom view of the $J-V$ and $P-V$ characteristics.  Polar plots of (e) $J_{sc}$ and PCE and (f) $V_{oc}$ and FF for the proposed HNW PSC for $\varphi = 0^\circ$ to $\varphi = 360^\circ$.}

    \label{Fig:fig8}
\end{figure*}

\begin{figure*}[h]
  \centering
   \includegraphics [width=1\textwidth]{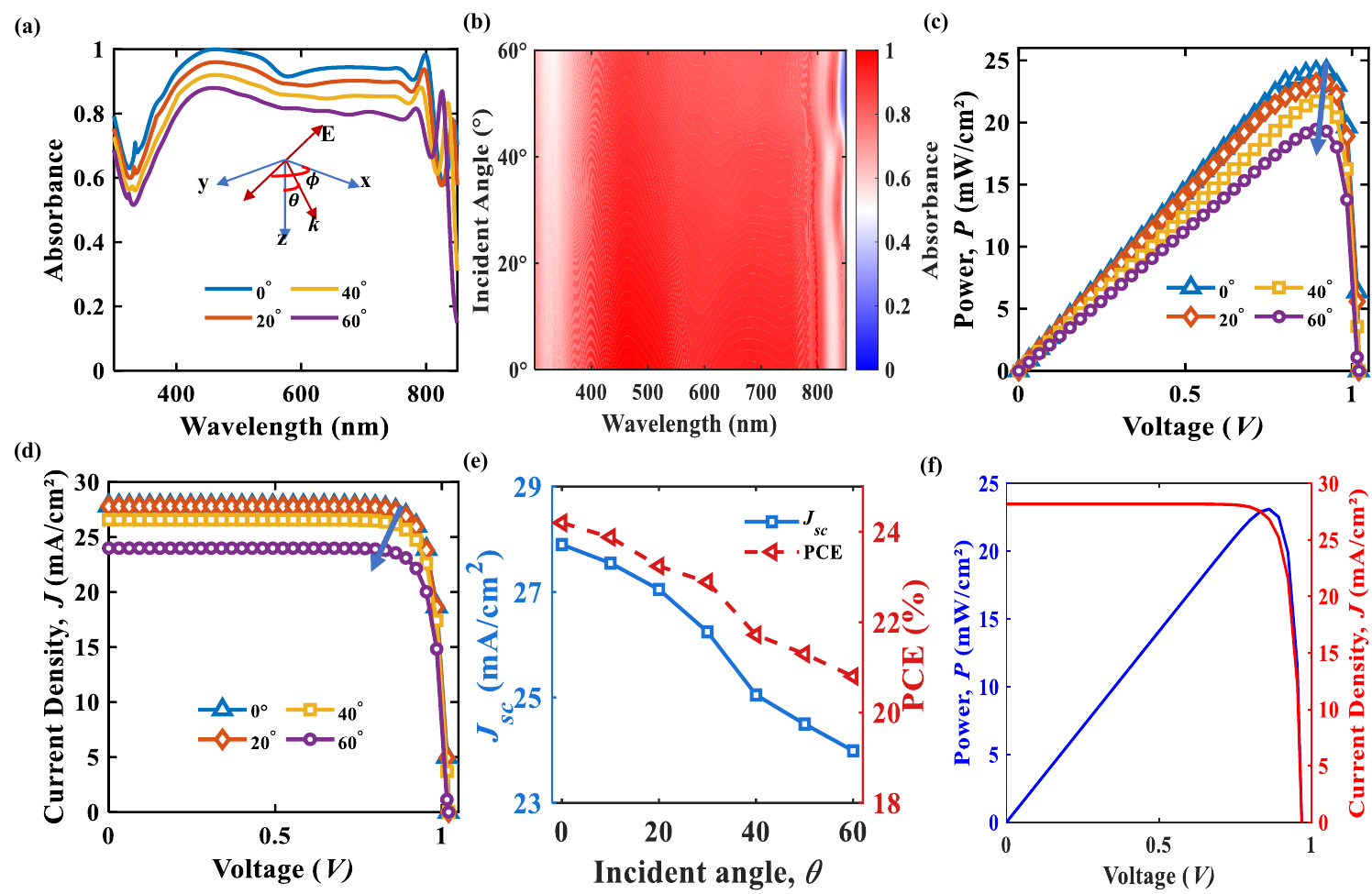}
    \caption{(a) and (b) Absorbance spectra at $\theta = 0^\circ$ to $\theta = 60^\circ$; the inset shows the polarization angle, $\varphi$, and the incident angle, $\theta$ of the plane wave, (c) and (d) $P-V$ and $J-V$ curve for unpolarized incident light at $\theta = 0^\circ$ to $\theta = 60^\circ$  for the proposed structure, (e) comparison of electrical parameters ($J_{sc}$, PCE) under unpolarized incident light at $theta = 0^\circ$ to $\theta = 60^\circ$, and (f) $J-V$ and $P-V$ for the proposed HNW PSC at normal incident light for unpolarized light source.}
    \label{Fig:fig9}
\end{figure*}

\subsection{Impact of Light Polarization on Device Performance}
To attain the outstanding PV performance, it is crucial to ensure polarization-independent absorbance capabilities across the entire wavelength range. This study examined both optical and electrical characteristics under polarized light illumination. Figs.\,\ref{Fig:fig8} (a) and (b) represent the line and contour plots of the absorbance spectra corresponding to the presence of various polarization angles ranging from $\varphi = 0^\circ$ to $\varphi = 90^\circ$. The absorbance spectra exhibit negligible variation across varying polarization angles, consistently displaying high absorbance levels ($>85\%$) throughout the wavelength range of 300-850~nm. This validates the spectral uniformity and indicates a stable absorbance band that remains largely unaffected by changes in polarization. Moreover, the average absorbance($(A_{avg})$) showed only a slight variation due to the variation in polarization angle from $\varphi = 0^\circ$ to $\varphi = 90^\circ$. The ($A_{avg}$) decreased from 89.86\% to 85.83\%, which denoted a relative change of 4.03\%. In addition, we evaluated the performance using the unpolarized light source, which is illustrated in Fig.\,\ref{Fig:fig8} (a), where the consistency in spectral behavior confirms the isotropic interaction with light regardless of the polarization.

The electrical performance of the proposed  HNW PSC was investigated by importing the generation rate profile extracted from the optical simulations into the solver to perform a self-consistent solution of Poisson’s equation and the drift–diffusion equations for electrons and holes under AM~1.5G solar illumination (100~mW/cm$^2$). The electrical properties of the materials used in the simulation are given in Table~\ref{tab:material_properties} \cite{Tabrizi2021}.
Fig.\,\ref{Fig:fig8} (c) shows the simulated current density–voltage ($J$–$V$) curves across different polarization angles, ranging from $0^\circ$ to $90^\circ$. The figure shows only minor changes, indicating that the HNW architecture preserves essentially stable charge-carrier dynamics regardless of polarization angles. Similarly, the corresponding power density–voltage ($P$–$V$) curves shown in Fig.\,\ref{Fig:fig8} (d) further confirm that the device's consistent energy conversion behavior across the polarization states.
The extracted values of $J_{\text{sc}}$,$V_{\text{oc}}$, fill factor (FF), and PCE at each polarization angle are summarized in Table \ref{tab:polarization_independence}. As observed, all key figures of merit remain stable, with no meaningful performance loss across the polarization angles. This behavior is primarily attributed to the geometric symmetry of the HNW and its isotropic light coupling, which together yield a uniform optical and electrical response regardless of the incident light's polarization. 
Figs.\,\ref{Fig:fig8} (e) and (f) demonstrate the PV performance parameters as a function of polarization angle varied from $\varphi = 0^\circ$ to $\varphi = 360^\circ$. Both figures exhibit flawless circular and symmetrical distributions over the full cycle rotation. The absence of significant angular modulation indicates that the generated photocurrent and the carrier transport mechanisms are not perturbed by the polarization-dependent effects, underscoring the intrinsic isotropy of the HNW structure. 

\begin{table}[H]
\centering
\caption{Electrical performance parameters of the proposed PSC as a function of polarization angles ($\varphi$) ranging from $0^\circ$ to $90^\circ$}
\begin{tabular}{cccccc}
\toprule
$\varphi$ & $J_{sc}$ (mA/cm$^2$) & PCE (\%) & $V_{oc}$ (V) & FF \\
\midrule
0$^\circ$  & 27.90 & 24.2 & 1.02 & 0.85 \\
30$^\circ$ & 27.86 & 24.15 & 1.02 & 0.85 \\
60$^\circ$ & 27.83 & 24.13  & 1.02 & 0.85 \\
90$^\circ$ & 27.87 & 24.16 & 1.02 & 0.85 \\
\bottomrule
\end{tabular}
\label{tab:polarization_independence}
\end{table}

\subsection{Impact of Incident Angle on Device Performance}
To evaluate the angular dependency of the proposed HNW PSC, both optical and electrical simulations were conducted under oblique illumination of the unpolarized incident light source. The angle of incidence ($\theta$) of unpolarized light was systematically varied from $0^\circ$ to $60^\circ$, while all structural and material parameters were kept unchanged. Figs.\,\ref{Fig:fig9} (a) and (b) illustrate the absorbance spectra in both line and contour formats for different $\theta$ values. The device exhibits a broad and spectrally stable absorbance profile throughout the visible range, indicating that the HNW architecture effectively preserves optical resonances under oblique illumination. However, a gradual reduction in absorbance intensity is observed with increasing $\theta$, which can be attributed to the reduced effective optical path length and weaker coupling of incident photons within the proposed structure.

The $A\textsubscript{avg}$ decreases from 89.82\% at normal incidence to 77.27\% at $\theta = 60^\circ$, corresponding to a relative reduction of only 12.55\%. The influence of $\theta$ on electrical performance is shown in Figs.\,\ref{Fig:fig9}(c) and (d) through the $J$--$V$ and $P$--$V$ characteristics. As $\theta$ increases, both the $J\textsubscript{sc}$ and maximum output power ($P_{\max}$) decrease in accordance with the reduced optical absorbance. Under normal incidence, $J\textsubscript{sc}$ and $P_{\max}$ reach values of 27.85 mA/cm$^{2}$ and 24.17 mW/cm$^2$, respectively. At $\theta$ = $20^\circ$, the average absorbance decreases, resulting in Jsc and Pmax of 27.04 mA/cm$^2$ and 23.24 mW/cm$^2$, respectively, corresponding to a relative change of 2.9\% and 3.97\% compared to the normal incidence. Over the full angular range from $0^\circ$ to $60^\circ$, the proposed structure maintains a $J\textsubscript{sc}$ of 27.85 and 23.99 mA\,cm$^{-2}$ while the $V\textsubscript{oc}$ and  FF remain nearly constant, indicating robust carrier transport characteristics. Fig.\,\ref{Fig:fig9} (e) compares the variation of $J\textsubscript{sc}$ and PCE as a function of $\theta$. The detailed PV parameters are summarized in Table \ref{tab:incident_angle_independence}.The resulting $J$–$V$ and $P$–$V$ characteristics under the normal incidence of unpolarized light are presented in Fig.\,\ref{Fig:fig9} (f), demonstrating strong current generation and high power output.

\begin{table}[H]
\centering
\caption{Electrical performance parameters of our proposed PSC for different incident angles $\theta$ ranging from $0^\circ$ to $60^\circ$}
\begin{tabular}{cccccc}
\toprule
$\theta$  & $J_{sc}$ (mA/cm$^2$) & PCE (\%) & $V_{oc}$ (V) & FF \\
\midrule
0$^\circ$  & 27.85 & 24.15 & 1.02 & 0.85 \\
20$^\circ$ & 27.04 & 23.44 & 1.02 & 0.85 \\
40$^\circ$ & 25.05 & 21.72  & 1.02 & 0.85 \\
60$^\circ$ & 23.99 & 20.8 & 1.02 & 0.85 \\
\bottomrule
\end{tabular}
\label{tab:incident_angle_independence}
\end{table}

\subsection{Proposed Cell Fabrication Method}
The device can be fabricated on cleaned glass substrates, where Al back electrodes can be thermally deposited in a high-vacuum environment to form the reflective bottom contact, following the typical approach for n-i-p PSCs \cite{Burschka2013}. After that, an HTL layer made of spiro-OMeTAD can be deposited via spin coating from a chlorobenzene solution containing Li-trifluoromethanesulfonylimide (TFSI) and tert-butylpyridine (tBP), followed by controlled drying under low-humidity conditions to promote oxidation and increase conductivity \cite{Jeon2014}. The HNW-based absorber layer made of MAPbI$_3$, which is intended to be fabricated utilizing an anodic aluminum oxide (AAO) template that inherently retains a self-structured hexagonal pore design generated during the electrochemical anodization of Al under fixed voltage and electrolyte conditions, and is laminated onto the spiro-OMeTAD/Al layer \cite{Xu2013}. The PbI$_2$ precursor solution can be infiltrated into the template's pores and converted to CH$_3$NH$_3$PbI$_3$ by immersing it in methylammonium iodide, followed by an annealing process. To achieve the desired heights of the NW, multiple infiltration-conversion cycles may be employed, followed by a partial removal of the template using dilute H$_3$PO$_4$ to reveal the wire tips while preserving the voids between the wires \cite{Chen2019}. To mitigate parasitic absorbance and moisture penetration in these void areas, insulating fillers like Polymethyl methacrylate (PMMA) can be added through spin coating from an anisole solution, due to its optical transparency ($k \approx 0$ in the visible spectrum), moderate refractive index ($n \approx 1.48$), and proven compatibility with perovskite devices \cite{Noel2014}. In addition, low-refractive-index materials like CYTOP fluoropolymer ($n \approx 1.34$) or MgF$_2$ ($n \approx 1.38$) can also be employed to infill the cavities while introducing negligible optical loss \cite{Brongersma2014}. After infilling, the NW sidewalls will undergo a short treatment with phenethylammonium iodide (PEAI) to passivate interfacial defects and enhance carrier lifetime \cite{Yang2015}. A compact ETL of TiO$_2$ will then be deposited through atomic layer deposition to guarantee close contact with the perovskite NW and facilitate efficient electron extraction \cite{Yang2015}. The final step involves forming a transparent conducting electrode by RF sputtering of ITO, thereby completing the vertical charge transport pathway. In order to enhance optical scattering and near field confinement, SiO$_2$ nanospheres synthesized via the Stöber method can be applied to the ITO using spin coating, followed by partial embedding through the deposition of an additional thin ITO layer. The hemispherical SiO$_2$ structures are anticipated to function as photonic scatterers to improve light coupling, aligning with prior strategies in nanophotonic light management for thin-film PVs \cite{Catchpole2008}.

\section{Comparative Analysis}
Table \ref{tab:comparison} demonstrates the comparative analysis of the PV performance of our proposed HNW PSC with the previously reported works. Fu \textit{et al.} presented a cylindrical MAPbI\textsubscript{3} microstructured cell that demonstrated improved optical confinement and achieved a PCE of about 20.19\%, with an average absorbance exceeding 90\% in the visible region, attributed to enhanced light absorbance facilitated by the microcavity design. Similarly, Joyti \textit{et al.} reported a 3D/3D MAPbI\textsubscript{3}/FAPbI\textsubscript{3} bilayer heterojunction that achieved an efficiency greater than 23\% while also showing enhanced device stability. In another study, Noori \textit{et al.} introduced a MAPbI\textsubscript{3}/CsPbI\textsubscript{3} heterojunction that achieved a PCE of 20.5\%, emphasizing the important role of band alignment in efficient charge separation and transport. Though the above-mentioned works mainly focused on optical confinement, some researchers placed their interest on modifications to the transport layer. Jing \textit{et al.} reported Sb\textsubscript{2}Se\textsubscript{3} as a narrow-bandgap HTL, improving hole mobility and reducing recombination losses, while Li \textit{et al.} conducted simulations on SnS\textsubscript{2}/MoS\textsubscript{2}-based ETL and HTL combinations to balance charge extraction. Likewise, Son \textit{et al.} further optimized the internal device parameters and reached an efficiency of 19.3\%, even if their planar structure was limited by optical path length and trapping capabilities. Notably, all the above studies are essentially focused on improving either optical absorbance or charge transport processes. However, few studies have incorporated both of these aspects into a single design. 

On the other hand, the HNW structure presented in this study incorporates both geometric light management and electrical optimization, resulting in a theoretical PCE of 24.2\% that outperforms all the above-mentioned studies. The NW geometries open up a new room for enhanced photon management and carrier transport. This improvement is due to the combined effects of Mie and FP-cavity while maintaining polarization robustness under TE, TM, and unpolarized illumination. Overall, the result highlights the  advantage of structural photonic design over conventional planar or bilayer configurations.

\begin{table*}[!t]
\centering
\caption{Comparison of performances among our work and previously reported works  }
\label{tab:comparison}
\setlength{\tabcolsep}{6pt}           
\renewcommand{\arraystretch}{1.15}    
\begin{tabular}{l l c c c c c}
\toprule
Structure & Active Layer & $\mathrm{J_{sc}}$ (mA\,cm$^{-2}$) & $\mathrm{V_{oc}}$ (V) & FF & PCE (\%) & Ref. \\
\midrule
HNW    & $\mathrm{CH_{3}NH_{3}PbI_{3}}$            & 27.90 & 1.02 & 0.85 & 24.2  & This work \\
Inverted Pyramid    & $\mathrm{CH_{3}NH_{3}PbI_{3}}$            & 27.81 & 0.95 & 0.82 & 21.85 & \cite{ULLAH2023170994} \\
Cylindrical & $\mathrm{CH_{3}NH_{3}PbI_{3}}$  & 24.50 & 0.97 & - & 20.19 & \cite{Fu} \\
Planar & $\mathrm{CH_{3}NH_{3}PbI_{3}}$/FAPbI$_3$  & 25.00 & 1.16 & 0.79 & 23.08 & \cite{Patil} \\
Planar & $\mathrm{CH_{3}NH_{3}PbI_{3}}$/CSPbI$_3$  & 24.66 & 1.06 & 0.78 & 20.5  & \cite{Noori2025} \\
Planar & $\mathrm{CH_{3}NH_{3}PbI_{3}}$  & 43.03 & 0.65 & 0.785 & 21.90 & \cite{Wu2025} \\
Planar & $\mathrm{CH_{3}NH_{3}PbI_{3}}$            & 24.91 & 0.97 & 0.74 & 19.30 & \cite{Son} \\
Planar & $\mathrm{CH_{3}NH_{3}PbI_{3}}$  & 25.72 & 0.84 & 0.78  & 16.95 & \cite{Li2024} \\

\bottomrule
\end{tabular}
\end{table*}

\section{Conclusion}
In this paper, we meticulously designed an ingenious HNW PSC structure and systematically elucidated the optoelectrical performance by integrating electromagnetic field analysis with carrier dynamics governed by the Poisson and drift-diffusion equations. In contrast to planar or cylindrical nanostructures, the rotational symmetry offered by the HNW structure enables broadband absorption in both TE and TM polarizations, ensuring polarization-independent absorbance across the entire wavelength spectrum. By partially inserting a dielectric SiO$_2$ sphere into the ITO layer, the design effectively utilized both Mie resonance and FP-cavity, which also enhanced near-field confinement, resulting in more efficient electron-hole generation near the active region. Detailed optical simulations showed that the proposed design greatly reduced reflectance and transmittance losses. Electrical simulations confirmed these improvements, yielding a high PCE of 24.2\% with corresponding FF and V\textsubscript{oc} of 0.85 and 1.02 V, respectively, under AM 1.5G solar illumination. Besides the quantitative results, this study highlights the important relationship among nanoscale geometry, light-matter interactions, and carrier dynamics. Ultimately, this study demonstrates that combining photonic confinement with electronic transport in the HNW geometry can enable the development of innovative, highly efficient PSCs using scalable and flexible design strategies.   

\printcredits
\section*{Data Availability Statement}
{The data supporting the findings presented in this paper are not currently available to the public, but they may be obtained from the authors upon reasonable request.} 

\section*{Acknowledgements}
The authors thank the Bangladesh University of Engineering and Technology (BUET) for providing technical support. 

\bibliographystyle{unsrt}

\bibliography{main}

\balance

\end{document}